\documentclass[11pt,twoside]{article}
\usepackage{asp2004}
\usepackage{psfig}
\usepackage{epsf}
\usepackage{graphics}
\usepackage{lscape}
\makeindex 
\begin{document}
\title{Understanding the Relationship Between Observations and Stellar Parameters in an Eclipsing Binary System}
\author{O.~L.~Creevey$^{1,2,3}$, 
Timothy M.~Brown$^{1}$, 
Sebastian Jim\'enez-Reyes$^{2}$\\ 
and
Juan Antonio Belmonte$^{2}$}

\affil{$^1$High Altitude Observatory, NCAR, Boulder 80307, CO USA}
\affil{$^2$Instituto de Astrof\'isica de Canarias, E-38200 La Laguna, Tenerife, Spain}
\affil{$^3$Universidad de La Laguna, 38206 La Laguna, Tenerife, Spain}

\begin{abstract}                                                              
We would like to investigate 
the information contained in our observations and to what extent  
each of them contributes individually to constraining the 
physical parameters of the system we are investigating.
To do this, we present a study involving the technique of 
Singular Value Decomposition
using as a simple example a detached eclipsing binary system.
We intend to apply an extension of
this technique to asteroseismic measurements of Delta~Scuti stars that are
members of eclipsing binary systems.
\end{abstract}

\section{Introduction\label{sec:intro}} 
Asteroseismology is the study of the oscillations of the stars.
It is the only method which 
allows us to {\it see} beyond the photosphere of a star, and learn 
about their interior conditions.  We are interested in $\delta$ Scuti stars, 
whose pulsation amplitudes are large enough to be detected from the Earth.  
However, they present a physical problem: not all of the oscillation 
modes that theory 
predicts are excited nor observed, which leads to various interpretations of 
the frequency spectra and thus a poorly-constrained solution 
(mass, age etc.). 
If we study $\delta$ Scuti stars in eclipsing binary systems we could 
accurately derive their stellar masses and radii.  Then we 
could constrain some of the parameters of the system.  This in turn would 
lead to an easier identification of the modes of the stars, allowing
us to better determine the real parameters of the system.
Combining information from these different sources
has been suggested by many authors, and some studies 
have been conducted with this in mind
\citep{aerts04,rodr04}.  
In fact, they agree that this is the only way 
to advance if we would like to test stellar structure and evolution theories.
\index{$\delta$~Scuti stars}

Following a technique similar to that conducted by \citet{brown94}, we would 
like to investigate how the combination of information  
from the study of binary systems with oscillation frequencies
constrains the solution parameters.  
In particular we would like to understand by how much each 
observable (radial velocities, oscillation modes etc.) 
influences each of the parameters of the system (mass, age etc).
To approach this problem we use the technique of Singular 
Value Decomposition (SVD) \citep{press} and apply it to a simple case of a 
detached eclipsing binary system.  
\index{Singular Value Decomposition, SVD}

\begin{table}[t]
\center{\caption{Initial parameters and observational errors}}
\label{table:param}
\footnotesize
\begin{center}
\begin{tabular}{lrclr}
\noalign{\smallskip}
\tableline
Parameters & $P_j$ &\hspace{1cm} & Observables & $\sigma_i$ \\
\noalign{\smallskip}
\tableline

\noalign{\smallskip}
$R_A$(R$_{\odot}$)  & 0.850 && $K_A$(kms$^{-1}$)  & 0.90\\
$R_B$(R$_{\odot}$) &  0.095 && $K_B$(kms$^{-1}$) &  0.90 \\
$i$($^{\circ}$) &  89.300&&D$_P$(mag) &  0.02 \\
$T_2/T_1$ & 0.300 && L$_F$(hrs) &  0.20\\
a(AU) &     0.060 && L$_T$(hrs) &  0.20\\
$M_A$(M$_{\odot}$) & 0.970&& P(yrs) & 5x10$^7$\\
$M_B$(M$_{\odot}$) & 0.100 && D$_S$/D$_P$ &0.012\\
\noalign{\smallskip}
\tableline
\end{tabular}
\end{center}
\end{table}

\section{Mathematical Background\label{sec:math}} \index{math}
For a more detailed discussion we refer the reader to \citet{brown94}.
Let $B_i, i=1,M$ be the observables of some system, and $P_j, j=1,N$ with $M>N$
be the variables (parameters) that define this system.
There exists some model or list of equations that when given
$P_j$, will result in $B_i$.  
In real situations we know $B_i$ and we would like to extract $P_j$.
If the system or model were linear, this
would be achieved relatively easy by
solving a system of M linear equations in N variables. 
However, normally this is not the case, and so by linearizing the system 
around a reference set of parameters, $P_{j0}$, we can write
\begin{equation}
B_{i} = B_{i0} + \sum^{N}_{j=1} \frac{\partial B_i}{\partial P_j} \delta P_j 
\label{eqn:bi}
\end{equation}
where $B_{i0}$ are the observables resulting from the initial parameter
estimates $P_{j0}$, 
$\delta P_j=P_j-P_{j0}$ 
and each of the derivatives, 
$\frac{\partial B_i}{\partial P_j}$ is evaluated at $P_j=P_{j0}$.
The parameter fitting problem 
involves finding $\delta P_j$. 
Given the real observables $\beta_i$ and after some substituting
our problem is mathematically equivalent to minimizing
$\chi^2 = || D \delta P - \delta B||^2$
where $\delta B_i = (\beta_i - B_{i0} ) /\sigma_i$,
and
\begin{equation}
D = \frac{1}{\sigma_i} \frac{\partial B_i}{\partial P_j}
\label{eqn:matrix}
\end{equation}

The solution is $\delta P = V W^{-1} U^T \delta B$,
where $D=UWV^T$ is the singular value decomposition of $D$.
It is precisely this decomposition that provides us with information 
regarding the origin of the errors in the parameter solutions. 
In the absence of the real observables $\beta_i$ we can 
still construct the derivative matrix $D$ to study the
transformation between the observable and the parameter space.

\section{Application to a Detached Eclipsing Binary System\label{sec:example}}
Given photometric and radial velocity measurements of a detached eclipsing
 spectroscopic binary system, we would like to obtain
 the following fundamental parameters: masses of both components $M_A$~and 
$M_B$, 
radii of both stars $R_A$~and $R_B$, effective temperature ratio $T_B$/$T_A$, 
inclination of system $i$ and separation of both components $a$.  
In order to obtain these parameters we need the following observables:
$K_A$~and $K_B$~are the semi-amplitudes of
the radial velocity curve.
From the photometric light curve we obtain the primary eclipse depth 
D$_P$ and with the secondary $D_S$, a depth ratio 
D$_S$/D$_P$.  We also observe a total duration of eclipse $L_T$,
and a duration of eclipse minimum, $L_F$ (see
assumptions below).
The photometric period $P$
is obtained from long term observations.
Let us also assume for simplicity that there is no eccentricity $e=0$, 
that neither of the components have stellar spots, 
that the secondary eclipse is total and 
that both stars are well isolated such that there
is no mass transfer between them.
\begin{figure}[t]
\centerline{\psfig{figure=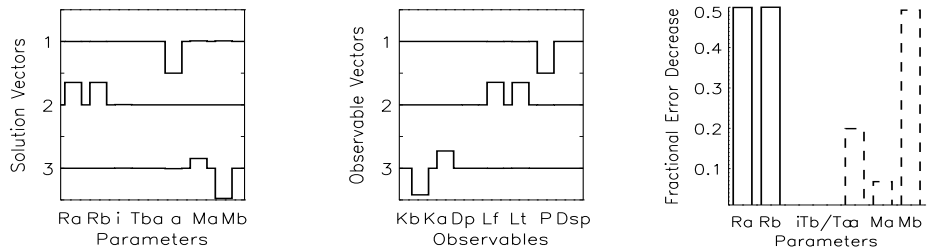,angle=0,clip=,height=3.8cm,width=13.5cm}}
\caption[]{
Information from the decomposition matrices of the derivative matrix $D$.
{\sl Left:} Three parameter solutions showing the extent to which the
parameters are constrained.   
{\sl Center:} Three observable vectors quantifying the amount of information
contributed by each observable.
{\sl Right:} The effect of increasing the precision of some observations by a 
factor of two on each of the solution parameter errors.}

\label{fig:fig1creevey1} 
\end{figure}

The {\it model} that relates these given observables to the parameters 
is a set of equations that can be easily derived.
After obtaining observations with errors $\sigma_i$, 
we make some initial parameter estimates
$P_{j0}$ (Table 1). 
We can then calculate the matrix $D$ (equation \ref{eqn:matrix}). 
Using SVD, we decompose $D$ to obtain its singular values and 
its corresponding orthogonal matrices. 
Figure \ref{fig:fig1creevey1} shows the first three solution vectors (left 
panel) and 
observable vectors (center panel)
of $D$.  The vector corresponding to the largest singular value (smallest
error)
is the top solution vector. 
According to the figure, the 
best-constrained parameter is the separation between both components
(left panel, top row vector).  There is also a small effect from
this singular value on both masses.
If we inspect the observable vector (center panel, 
top row
vector), 
we can see clearly that $P$ is
the most reliable observation (corresponding to the largest 
singular value).  
We know that $P$ is directly related to $a$ and $M_A+M_B$
through Kepler's Law giving a well-determined $a$. 
In this case $M_A$ and $M_B$ are mainly constrained 
by the radial velocity measurements (left and center panel, bottom row vector).

The second solution vector shows that both 
$L_T$ and $L_F$ are also some of the more important observables
in constraining the solution.  Figure \ref{fig:fig1creevey1} left panel
shows that both $R_A$ and $R_B$ correspond to this singular value,
implying that $L_F$ and $L_T$ are mainly responsible for the reduction in the
errors of both radii.
In fact, $L_F$ and 
$L_T$ contribute also to $i$ but to a lesser extent (sixth singular value).


Suppose now that the errors in our measurements ($\sigma_{i}$) were 
to decrease by a factor of 2.   By decreasing each of the $\sigma_{i}$ 
individually, we may investigate how each of the errors on the 
parameters ($\epsilon_{j}$) improves (noting that combining all $\sigma_{i}$ 
should result in a corresponding reduction of 0.5 in all the $\epsilon_{j}$).  
Figure 1 right panel indicates the decrease in each $\epsilon_{j}$  
as a fraction of the original error 
($\epsilon_{{j0}} - \epsilon_{{ji}})/ \epsilon_{{j0}} $, where 
$\epsilon_{{j}}$ comes from the covariance matrix (Press et al. 1986), 
and 
$\epsilon_{{j0}}$ represents $\epsilon_{{j}}$ given the original 
$\sigma_{{i0}}$, and 
$\epsilon_{{ji}}$ are the resulting $\epsilon_{{j}}$ given a decrease 
in a particular $\sigma_{i}$.
We show the effects of reducing just $\sigma_{D_P}$ (solid) and 
$\sigma_{K_A}$ (dash).  

The new $\sigma_{D_P}$ is solely responsible for the 
reduction of both  $\epsilon_{R_A}$ and $\epsilon_{R_B}$ (a total of 0.5). If 
we were interested in improving the errors on both radii, then we know from 
this that we need only improve the precision of $D_P$ (note with $\sigma_{i0}$ 
$L_F$ and $L_T$ were important for $R_A$ and $R_B$).
Similarly, $\sigma_{K_A}$ is responsible for all of the reduction of 
$\epsilon_{M_B}$.  However, $\sigma_{K_A}$ is also {\it partially}
 responsible for reducing both $\epsilon_{M_A}$ and $\epsilon_{a}$ (by 0.2 and
0.1 respectively).  
This indicates that if we were interested in improving 
the precision of $M_A$, we would need not only an improvement of 
$\sigma_{K_A}$, but a combination of improved $\sigma_{i}$ (in this case
$\sigma_{K_B}$, $\sigma_{L_F}$ and $\sigma_{L_T}$).
We should also take into account the actual values of the errors, for example,
if $\epsilon_{M_A}$ improves from 0.004 to 0.002 $M_{\odot}$ 
(which is this case), 
we could also conclude that these observations may not be worthwhile.
So we must look at these results cautiously, and we must also 
emphasize the sensitivity of these results to $\sigma_{i0}$.

If we could decrease each $\sigma_{i}$ by 3, 4, 5... times,
we could determine at what error level this observable becomes important
in constraining a particular parameter, given all the other $\sigma_i$.
This could indicate if it were worthwhile
obtaining better observations.  It may also help determine the
precision of our observations necessary to obtain
a solution to the system with errors to within a certain \%.
Studying the transformation
between parameter and observable space may allow us to reach these
conclusions.

\section{Conclusions\label{sec:conclusions}}
What we have discussed is just a simple example to illustrate 
what we would 
like to achieve in the case of a more complicated system, such as a detached
eclipsing binary system where one of the components is a $\delta$ Scuti star.
Unfortunately, for that case we can not write simple equations to relate the observables
to the parameters, and for this reason a study such as this one is 
beneficial.

We saw briefly the capacity of SVD to try to explain the relationship 
between the errors in each of the observables to the errors in each of 
the parameters.  We saw how the effect of an increase in the 
precision of the 
observables can
impact in a rather complex form on the resulting errors of the parameters.  

Ideally we would like to observe a real eclipsing system with a $\delta$
Scuti component.  From studying this transformation we hope to be able to 
optimise our observations to obtain the best constrained parameters.
Doing this would allow us to {\it solve} this system to the required 
precision in order 
to test theories of stellar structure.  

{}

\end{document}